\pdfoutput=1
\documentclass[format=acmsmall, review=false, screen=true]{acmart}
\usepackage{booktabs} 

\makeatletter
\renewcommand\footnotetextcopyrightpermission[1]{}
\makeatother
\usepackage[ruled]{algorithm2e} 

\SetAlFnt{\small}
\SetAlCapFnt{\small}
\SetAlCapNameFnt{\small}
\SetAlCapHSkip{0pt}
\IncMargin{-\parindent}

\usepackage[normalem]{ulem}
\usepackage{arydshln}
\usepackage{pbox}
\usepackage{amsmath}
\usepackage{amsfonts}
\usepackage{amssymb}
\usepackage{url}
\usepackage{comment}
\usepackage{graphicx}
\usepackage{epstopdf}
\usepackage{multirow}
\usepackage{color}
\usepackage{mdframed}
\usepackage{syntax}
\usepackage{algpseudocode}
\usepackage{stmaryrd}
\usepackage{proof}
\usepackage{tikz}
\usepackage[referable]{threeparttablex}
\usepackage{paralist}
\usepackage{listings}

\usetikzlibrary{calc}

\newcommand{\fixme}[1]{\footnote{\textbf{\color{red}{FIXME:}} #1}}

\providecommand{\bd}[0]{\mathbb{B}}

\setcopyright{none}

\begin{document}
\title[Instruction-Level Abstraction for SoC Verification]
{Instruction-Level Abstraction (ILA): A Uniform Specification for System-on-Chip (SoC) Verification}

\author{Bo-Yuan Huang}
\email{byhuang@princeton.edu}
\affiliation{%
  \institution{Princeton University}
  \city{Princeton}
  \state{NJ}
  \postcode{08544}
  \country{USA}
}

\author{Hongce Zhang}
\email{hongcez@princeton.edu}
\affiliation{%
  \institution{Princeton University}
  \city{Princeton}
  \state{NJ}
  \postcode{08544}
  \country{USA}
}

\author{Pramod Subramanyan}
\email{pramod@berkeley.edu}
\affiliation{%
  \institution{University of California, Berkeley}
  \city{Berkeley}
  \state{CA}
  \postcode{94720}
  \country{USA}
}

\author{Yakir Vizel}
\email{yakir.vizel@gmail.com}
\affiliation{%
  \institution{Princeton University}
  \city{Princeton}
  \state{NJ}
  \postcode{08544}
  \country{USA}
}

\author{Aarti Gupta}
\email{aartig@cs.princeton.edu}
\affiliation{%
  \institution{Princeton University}
  \city{Princeton}
  \state{NJ}
  \postcode{08544}
  \country{USA}
}

\author{Sharad Malik}
\email{sharad@princeton.edu}
\affiliation{%
  \institution{Princeton University}
  \city{Princeton}
  \state{NJ}
  \postcode{08544}
  \country{USA}
}

%
%

%
%


\begin{abstract}

Modern Systems-on-Chip (SoC) designs are increasingly heterogeneous and contain specialized semi-programmable accelerators in addition to programmable processors. In contrast to the pre-accelerator era, when the ISA played an important role in verification by enabling a clean separation of concerns between software and hardware, verification of these ``accelerator-rich'' SoCs presents new challenges. From the perspective of hardware designers, there is a lack of a common framework for formal functional specification of accelerator behavior. From the perspective of software developers, there exists no unified framework for reasoning about software/hardware interactions of programs that interact with accelerators. 

This paper addresses these challenges by providing a formal specification and high-level abstraction for accelerator functional behavior. It formalizes the concept of an Instruction Level Abstraction (ILA), developed informally in our previous work, and shows its application in modeling and verification of accelerators. This formal ILA extends the familiar notion of instructions to accelerators and provides a uniform, modular, and hierarchical abstraction for modeling software-visible behavior of both accelerators and programmable processors. We demonstrate the applicability of the ILA through several case studies of accelerators (for image processing, machine learning and cryptography), and a general-purpose processor (RISC-V). We show how the ILA model facilitates equivalence checking between two ILAs, and between an ILA and its hardware finite-state machine (FSM) implementation. Further, this equivalence checking supports accelerator upgrades using the notion of ILA compatibility, similar to processor upgrades using ISA compatibility.

\end{abstract}

\maketitle

\renewcommand{\shortauthors}{B.-Y. Huang, H. Zhang, P. Subramanyan, Y. Vizel, A. Gupta, S. Malik}

\section{Introduction}
\label{sec:intro}

Today's computing platforms are increasingly heterogeneous, a trend that is expected to continue into the foreseeable future as per the International Technology Roadmap for Semiconductors~\cite{chan2014itrs}. In addition to programmable processors -- both general purpose and domain specific such as Graphics Processing Units (GPUs) -- today's platforms contain dedicated accelerators in order to meet the power-performance requirements posed by emerging applications. These accelerators may be tightly coupled, i.e., part of the processor pipeline, or loosely coupled, interacting with the processor through shared memory~\cite{cota2015analysis}. The latter form is dominant and the focus of this paper. Apple's {\tt A} series of processors illustrate this growth in accelerators; the {\tt A8} processor has 30 accelerators~\cite{shao2015-toppicks}  while the {\tt A10} has 40. 

Accelerator-rich platforms pose two distinct verification challenges. The first challenge is constructing meaningful specifications for accelerators that describe behavior exposed at the hardware/software interface. Such specifications are important not just for correct design/verification of hardware, but are also required to drive software and firmware development, both of which must often be done before the hardware is ``taped-out.'' Specifications are also required to reason about portability between different generations of accelerator architectures. They can mitigate the software incompatibility risk involved in the implementation of microarchitectural enhancements. Further, it is important to note that specifications must necessarily be an \emph{abstraction} of hardware functionality. Detailed models, e.g., Register-Transfer Level (RTL) descriptions, expose cycle-level behavior that is not part of the hardware/software interface and thus are not suitable as specifications. In addition, RTL descriptions are also undesirable as specifications because the detailed nature of these models means they are not amenable to scalable formal analysis. 



The second challenge is reasoning about hardware-software interactions from the perspective of software. For software that runs exclusively on a programmable processor, its execution semantics are defined by the processor's instruction set architecture (ISA) specification. Thus, the ISA serves as a suitable abstraction of the underlying processor hardware for software verification. 
However, similar abstractions of hardware for reasoning about software interacting with accelerators are lacking. Software typically accesses accelerators through memory-mapped input-output (MMIO) instructions that map memory and registers inside the accelerators to specific addresses. From the perspective of the ISA, accelerator interactions appear to be just loads/stores of these addresses. However these loads/stores trigger specific functionality implemented by the accelerator logic not modeled by the processor's load/store instruction semantics. Further, the accelerator may access some memory shared with the processor, and potentially interrupt the processor on completion of specific functions. These aspects make the ISA incomplete for modeling accelerator interactions. As a result, reasoning about software that interacts with accelerators, an increasingly important task in today's SoCs, is usually done through ad-hoc abstractions/modeling techniques that compose ISA-level models with FSM models of accelerators (e.g., in Verilog/VHDL). This results in an \emph{abstraction gap} between the ISA and the low-level hardware FSM, making software/hardware co-verification with accelerators very challenging.  


In this work, we propose a uniform and formal abstraction for processors and accelerators that captures their software-visible functionality. This abstraction is called an Instruction-Level Abstraction (ILA) and is based on the familiar notion of computation triggered by ``instructions.'' For a processor, the ILA is based on the ISA. For an accelerator, the insight is that commands at its interface are akin to instructions in a processor. Thus, just as the ISA models processor behavior through specifying state changes resulting from each instruction, the ILA models accelerator behavior by specifying state changes resulting from each of its ``instructions,'' i.e., its  commands. Further, as with ISAs, this modeling can distinguish the state that is persistent between instructions (architectural state), from implementation state (micro-architectural state). Top-down this modeling provides a specification for functional verification of hardware, and bottom-up it provides an abstraction for software/hardware co-verification. 

The ILA, like an ISA, has the following useful attributes. It provides: 
\begin{enumerate}[(i)]
    \item a modular functional specification as a set of instructions;
    \item a meaningful state abstraction in terms of architectural state, \emph{i.e.}, state that is persistent between instructions, while abstracting away implementation state; and
    \item a specification for each instruction in the form of state update functions for architectural state.
\end{enumerate}

In modeling designs with complex instructions, it is sometimes easier to describe the architectural state update function as a sequence of steps, i.e., an algorithm. These steps may be required of all implementations, in which case they are considered part of the specification, or may only indicate a possible implementation. The ILA model allows this sequencing to be expressed through hierarchy in instructions, where an instruction can itself be modeled as a \emph{sequence} of two different kinds of \emph{child} instructions.

This work builds on~\cite{subramanyan2015template, subramanyan2017template} which introduced an informal notion of the Instruction-Level Abstraction (ILA).
That work viewed an ILA as a finite state system and focused on synthesizing ILAs using program synthesis techniques~\cite{Sygus13,jha2010oracle}. 
The focus of this work is on formalizing the ILA as an instruction-centric operational model, well-suited as an interface between sequential software and the underlying hardware. To treat processors and accelerators uniformly, the ILA model explicitly includes functions that perform the fetch-decode-execute of instructions. This is especially useful in reasoning about a system of interacting ILA models, one ILA 
per processing unit, where the decode function (dependent on the fetch function) captures the condition whether an instruction is enabled to execute or not, and the execute part actually performs the update of the software-visible state. Note that the earlier finite state model could capture only the execute part. 
Furthermore, we have introduced hierarchy into the ILA model, via the notions of child (sub- and micro-) instructions, where an instruction at a higher level can be represented as a sequence of child instructions at a lower level. Thus, the granularity of ILA instructions can vary, ranging from processor instructions to software functions, but the focus is on modeling software-visible states and their updates. 
Finally, this work showcases the usefulness of the formal ILA model and its applications in verification through a set of rich case studies comprising accelerators from diverse applications domains (advanced encryption, image processing, machine learning) and a processor (RISC-V Rocket Core). The earlier papers had focused only on an accelerator for encryption. 

Note that while we describe the verification applications using ILAs in detail, we do not claim the verification techniques to be our central contribution -- indeed, we have used standard verification techniques and commercial off-the-shelf verification tools in our case studies. The point to note is that the ILA model enables application of these techniques in a compositional manner, where the set of instructions naturally provides an instruction-based decomposition into simpler verification tasks. 

\subsection*{Contributions of this Paper}
Overall this paper makes the following contributions:

\begin{itemize}
    \item It provides a formal model for the ILA (\S~\ref{sec-formal-def}). This addresses critical modeling issues in both processors and accelerators including gaps in previous ISA formal models. Top-down this model provides a formal specification for use in hardware verification, and bottom-up an abstraction for use in software/hardware co-verification that is \emph{uniform} across accelerators and processors. 
    \item It supports hierarchy (\S~\ref{sec:hierarchy}) in modeling instructions which is missing from the earlier formal ISA models~\cite{arm16cav}. In particular, it makes the important distinction between hierarchy in the specification and hierarchy in the implementation.
    \item It demonstrates the applicability of the ILA model through several case studies on accelerators (AES, RBM, Gaussian Blur) and the RISC-V Rocket processor (\S~\ref{sec:case}).
    \item It demonstrates the value in verification across models -- between two ILAs, and between ILA and FSM models -- through successful case studies (\S~\ref{sec-4}), including finding a bug in the RISC-V Rocket processor core. Verifying FSM implementations against ILA specifications provides the basis for ILA-compatible accelerator replacement.
\end{itemize}


\section{Motivation and Background} 
\label{sec:background}
\begin{table*}
\footnotesize
  \begin{center}
    \begin{threeparttable}
      \caption{Comparison of Hardware and System-Level Modeling Frameworks}
      \label{tab:modeling-frameworks}
      \begin{tabular}{lcccccc}
        \toprule
        \multirow{2}{*}{\bf Modeling Language/Framework} & \multicolumn{5}{c}{\bf Level of Abstraction} & \multirow{2}{*}{\parbox{2cm}{\centering{\bf Formal Semantics}}}\\
        \cmidrule{2-6}
        & {Alg.} & {Func.} & {CA} & {RTL} & {GL} \\
        
        \midrule
        Verilog/VHDL & & & \checkmark & \checkmark & \checkmark  & Yes \\
        Design Specific Models in C/C++ etc. (e.g.,~\cite{gem5-11, esesc-hpca-13, dramsim-11}) & \checkmark & \checkmark & \checkmark & & & No \\ 
        Chisel, PyMTL~\cite{chisel-12, lockhart-micro-14} & & \checkmark & \checkmark & \checkmark &  & No\\
        System Level Modeling Frameworks~\cite{panda-systemc-01,hoe2000synthesis,DBLP:conf/iccad/BerryKS03,nikhil-bluespec-04,DBLP:journals/tosem/HarelN96,DBLP:conf/dac/BacchiniGMKDMGN07,shen1999using}  & \checkmark & \checkmark & & & & Yes \\
        \midrule
        {ILA (this work)} & & \checkmark & \checkmark & \checkmark & & Yes \\
        \bottomrule
      \end{tabular}    
\begin{tablenotes}
\item Column labels are Algorithmic (Alg.), Functional (Func.), Cycle Accurate (CA), Register Transfer Level (RTL) and Gate Level (GL).
\end{tablenotes}
    \end{threeparttable}
  \end{center}
\end{table*} 
\subsection{System-Level/Hardware Modeling Frameworks}

Table~\ref{tab:modeling-frameworks} categorizes notable system-level and hardware modeling frameworks in terms of their level of abstraction and the suitability of their models for formal analysis. The ``traditional'' approach to processor-based platform design uses: (i) functional models of processor ISAs (typically developed in C/C++) to define architectural behavior, and (ii) cycle-accurate simulators (e.g., ESEC and gem5~\cite{esesc-hpca-13, gem5-11}, also in C/C++) to explore the microarchitectural design space. Finally, the implementation typically uses register-transfer level (RTL) descriptions in Verilog/VHDL. This approach corresponds to the first two rows in Table~\ref{tab:modeling-frameworks}.

Recent years have seen increased interest in system-level modeling that raises the level of abstraction for design and verification. SystemC in particular, has seen significant adoption in system/transaction-level modeling. However, RTL designs in Verilog, corresponding to SystemC transaction-level models, are usually separately constructed by hand. Ensuring that the system-level models in SystemC and the corresponding RTL are in agreement is a challenging problem. Chisel~\cite{chisel-12} and PyMTL~\cite{lockhart-micro-14} propose to address this challenge by providing unified domain-specific embedded languages in Scala and Python, respectively, for constructing functional, cycle-accurate, and RTL models. While this can mitigate some challenges in testing equivalence among these various models, bugs still slip through the cracks. In particular, these languages do not have formal precisely-defined semantics which limits automated reasoning. This makes it hard to provide guarantees of equivalence between models at different levels of abstraction.

Models with formally-defined operational semantics are amenable to formal analyses such as equivalence and property checking. 
Examples include StateCharts, SystemC, Esterel, Transaction Level Modeling (TLM), and others~\cite{DBLP:journals/tosem/HarelN96,DBLP:conf/iccad/BerryKS03, DBLP:journals/toplas/AlurG04, DBLP:journals/ijpp/AbdiG06, DBLP:conf/dac/BacchiniGMKDMGN07, DBLP:journals/ejes/DomerGPSCYAG08,DBLP:journals/tecs/HerberG13,hoe2000synthesis}. 
A notable effort in this category is BlueSpec, a high-level specification and design language that describes hardware as sets of state change rules (guarded atomic actions) which execute atomically~\cite{nikhil-bluespec-04,shen1999using}.
The BlueSpec compiler synthesizes the circuits and exploits parallelism with a scheduler to choose the interleaving of rules automatically~\cite{hoe2000synthesis,dave2007scheduling}.
BlueSpec has well-defined operational semantics and supports modular verification using SMT solvers and interactive theorem-provers~\cite{dave2011verification,vijayaraghavan2015modular}.\footnote{See \S \ref{sec:related} for a detailed comparison of the ILA with BlueSpec and other related efforts.}

\subsection{Desired Hardware Abstraction Characteristics}
A given hardware design can be abstracted in many different ways. In this paper, we argue for abstractions of hardware that satisfy two important properties: 
\begin{itemize}
    \setlength\itemsep{0em}
    \item The abstraction cleanly separates hardware and software verification concerns. This requires that the abstraction precisely codify the hardware/software interface so that software and hardware can be separately developed and verified to be conformant with the interface. 
    \item The abstraction treats programmable processors and accelerators uniformly. Software verification in future architectures will need to reason about accelerator interactions in addition to processor ISAs, while hardware verification will need to reason about the software interface presented by these accelerators. A uniform abstraction for these architectures is required in order to provide a common accelerator-agnostic framework for this verification.
\end{itemize}

None of the frameworks in Table~\ref{tab:modeling-frameworks} satisfy these properties. In this paper, we take a step towards addressing this gap by introducing a uniform and hierarchical instruction-level abstraction (ILA): an abstraction of hardware that precisely delineates the hardware/software interface. Our notion of the ILA treats programmable processors and semi-programmable accelerators uniformly, including hierarchical modeling of microarchitecture for accelerators, similar to processors. 
Past work has shown how abstractions at the instruction-level can be successfully used for software/hardware co-verification~\cite{subramanyan2016verifying}.

\section{Formal Modeling} \label{sec-formal-def}

\begin{figure*}
\centering
\includegraphics[width=\textwidth]{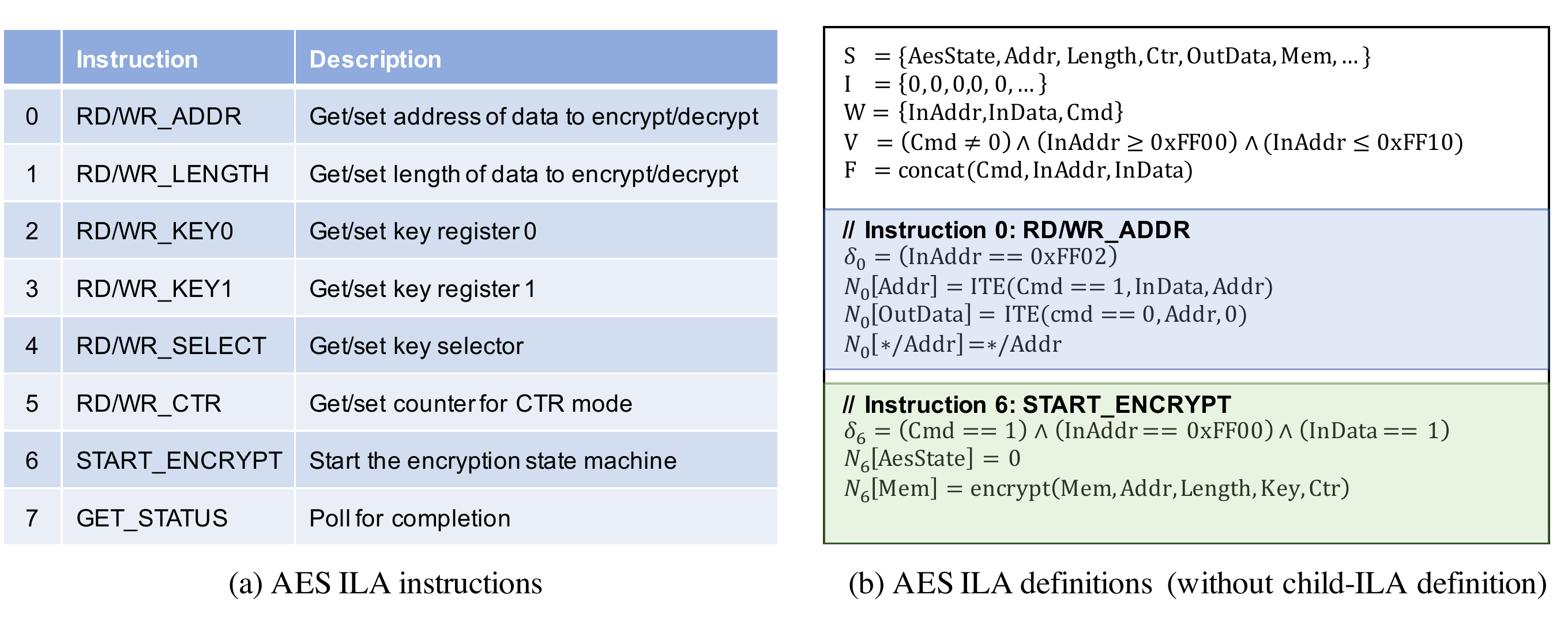}
\caption{ \label{fig:aes-ila} ILA for an AES accelerator.}
\end{figure*}

In this section, we formally define the ILA model and its execution semantics. A motivating example used through this section is shown in Figure~\ref{fig:aes-ila}, for an accelerator (from opencores.org)~\cite{ocAES} that implements the Advanced Encryption Standard (AES). 
The derived ILA instructions are shown in Figure~\ref{fig:aes-ila}(a): six instructions read/write configuration registers, one starts encryption, and one checks the completion status.
As discussed earlier, these ``instructions'' correspond to commands presented at the accelerator interface by the processor.


\subsection{Instruction-Level Abstraction (ILA)}
\label{sec:ilamodel}

\noindent
This section defines the ILA, without considering hierarchy. An ILA ${A}$ is a tuple: $\langle S, I, W, V, F, D, N \rangle$, where
\begin{eqnarray}
      &S& \text{ is a vector of state variables, } \nonumber \\
      &I& \text{ is a vector of initial values of the state variables, } \nonumber \\
      &W& \text{ is a vector of input variables, } \nonumber \\
      &V& : (S\times W) \to \bd \text{ is the valid function, } \bd = \{0,1\}  \nonumber \\
      &F& : (S\times W) \to bvec_w \text{ is the fetch function,} \nonumber \\
      &D& = \{ \delta_i: bvec_w \to \bd \} 
            \text{ is the set of decode functions,} \nonumber \\
      &N& = \{ N_i : (S \times W) \to S \} 
            \text{ are the next state functions.} \nonumber
\end{eqnarray}

The state variables in $S$ can be of type Boolean, bitvector, or array (representing memory). For processors, $S$  includes architectural registers, flag bits, data and program memory. For accelerators, $S$  includes memory-mapped registers, internal buffers, output ports to on-chip interconnect, data memory, etc. We refer to these state variables as ``architectural state'' because like an ISA's architectural state, they are persistent across instructions. 
In the ILA for the AES example, as shown in Figure~\ref{fig:aes-ila}(b), the architectural state variable $\texttt{Addr}$ denotes the address of data to encrypt, and $\texttt{Length}$ is the data length. 
$I$ denotes the set of initial values of the corresponding architectural states in $S$. 
The vector of input variables $W$ includes input ports of the hardware module, such as processor interrupt signals and accelerator command inputs. For example, input $\texttt{InData}$ in the AES ILA is the data from the memory system for memory-mapped accesses.

Instructions in an ILA follow the fetch/decode/execute paradigm, similar to a processor ISA. 
To model event-driven accelerators, we include a valid function $V: (S \times W) \to \bd$ that indicates if an instruction is triggered based on state and input values. For example, the AES accelerator executes instructions only when $\texttt{InAddr}$ is within a specified range, i.e., $V(S, W) \triangleq (\texttt{InAddr} \geq \texttt{\small 0xFF00}) \wedge (\texttt{InAddr} \leq \texttt{\small 0xFF10})$. 

The \textit{opcode} of the instruction is modeled as a bitvector of width $w$ (denoted $bvec_w$). If the instruction is triggered (\emph{i.e.}, if $V$ is true), then the fetch function $F:(S \times W) \to bvec_w$ indicates how it is extracted from the state and inputs. For processors, the opcode is fetched from the program memory location pointed to by the program counter, i.e., $F(S, W) \triangleq read( \texttt{IMEM}, \texttt{PC} )$. If interrupt modeling is desired, $F$ concatenates this with the interrupt signals (inputs).
Similarly, accelerators extract the opcode for decoding instructions. The opcode for the AES example is the concatenation of the memory-mapped input signals, as shown in Figure~\ref{fig:aes-ila}(b).

Each instruction (indexed by $i$) is associated with a decode function $\delta_i: bvec_w \to \bd$,  indicating whether it is \textit{issued}. 
For example, as shown in Figure~\ref{fig:aes-ila}(b), the instruction \texttt{\small START\_ENCRYPT} is issued only when it receives a ``store value 1 to address 0xFF00'' command at the interface.
The set of all decode functions is $D = \{ \delta_i ~|~ 0 \leq i < k \}$; $k$ is the number of instructions.
In an ILA, only one instruction can be issued at a time, i.e., $D$ is one-hot encoded. 
Non-determinism should be modeled with explicit \emph{choice variables} (inputs) provided by the external environment. 
Note the valid function $V$ returns true if and only if one decode function returns true. 

Finally, each instruction is associated with a next state function $N_i:(S \times W) \to S$, which represents the state update when the instruction is executed. 
The set of all next state functions in the ILA is $N = \{ N_i ~|~ 0 \leq i < k \}$. 

To summarize, Figure~\ref{fig:aes-ila} Part (a) shows the description of all eight instructions of the AES accelerator. Part (b) shows the ILA definitions for $S$, $I$, $W$, $V$, $F$, and the decode ($\delta_i$) and state update functions ($N_i$) for two of the instructions. 


\subsection{Hierarchical ILAs}
\label{sec:hierarchy}

\begin{figure*}
\centering
\includegraphics[width=\textwidth]{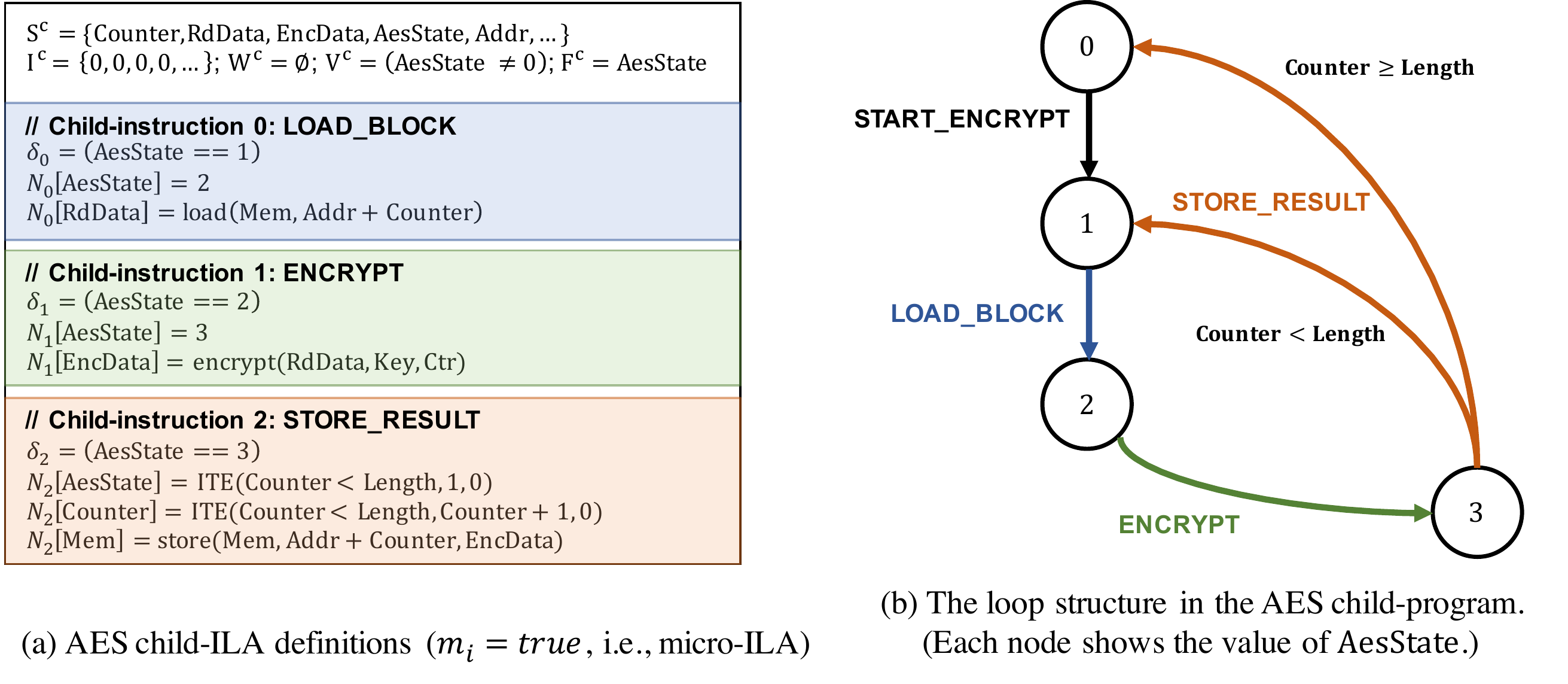}
\caption{ Child-ILA for an AES accelerator. }
\label{fig:aes-child-ila}
\end{figure*}

In modeling designs with complex instructions, it is often easier to describe the architectural state update function as a sequence of steps, i.e., an algorithm. These steps may be required of all implementations, in which case they are considered part of the specification, or may only indicate one possible implementation.
For example, the Intel x86 architecture~\cite{intel-manual} specifies the string copy instruction \texttt{\small REP MOVS} as a sequence where the \texttt{\small MOVS} instruction is repeated until register \texttt{\small ECX} (count) is decremented to $0$. 
Note that the state update performed by this instruction at the architectural level in not atomic, and this fact needs to be captured in the architecture model.
Similarly, in the AES accelerator in Figure~\ref{fig:aes-ila}, the \texttt{\small START\_ENCRYPT} instruction involves reading data, encrypting it, and writing the result. The encryption itself is also a complex operation that needs to be described as a sequence of steps.
\subsubsection*{Child ILAs}
To support modeling such complex instructions, we extend the ILA definition from Section~\ref{sec:ilamodel} to support hierarchy.
A hierarchical ILA may contain \textit{child-ILAs}, each of which describes the \emph{sequence of steps} in the complex instruction. 
Instructions in child-ILAs, referred to as child-instructions, also follow the fetch/decode/execute paradigm. 
These may, in turn, contain other child-ILAs, and we refer to an ILA containing a child as a \textit{parent-ILA}. 
In the AES example, \texttt{\small START\_ENCRYPT} is modeled by a child-ILA with child-instructions for message loading, encryption, and storing results, as shown in Figure~\ref{fig:aes-child-ila}~(a). 
The child-instruction \texttt{\small{ENCRYPT}} is further modeled by a child-ILA for the actual encryption algorithm. 
As we will show later in Section~\ref{sec-case-acc}, two different child-ILAs can be used to describe two different AES implementations.



A child-ILA is defined similar to an ILA. Its state variables are denoted $S^c$, some of which may be shared with the parent-ILA. $W^c$ is the set of its input variables, and is a subset of inputs of the parent-ILA. The initial values are $I^c$, and the initial values of the shared state variables are the same as that of the parent-ILA.
For the AES example in Figure~\ref{fig:aes-child-ila}~(a), the child-ILA has no inputs and contains three additional state variables (\textit{Counter}, \textit{RdData}, and \textit{EncData}). 
The other components of the child-ILA: $V^c$, $F^c$, $D^c$, $N^c$ are similarly defined in terms of $S^c$ and $W^c$. The state variables shared between the child-ILA and the parent-ILA are \emph{in lock step}, since they denote shared state, i.e., updates to the shared states are visible to the parent-ILA when a child-instruction is executed, and vice versa. 
For the AES example in Figure~\ref{fig:aes-ila}, the instruction \texttt{\small START\_ENCRYPT} of its parent-ILA updates shared state \texttt{AesState} to $1$ and keeps \texttt{Mem} unchanged; this starts the child-ILA with the child-instruction \texttt{\small LOAD\_BLOCK}.

Informally, a child-ILA steps through a sequence of child-instructions, where the sequencing is implicitly determined by the state updates (using its state variables). That is, the child-instructions are sequentially composed. This can be viewed as a \textit{child-program}. 
For the AES example, the child-program in Figure~\ref{fig:aes-child-ila}~(a) models the \texttt{\small{START_ENCRYPT}} instruction, which comprises a loop and is controlled by the states \texttt{AesState} and \texttt{Counter}, as illustrated in Figure~\ref{fig:aes-child-ila}~(b).



\subsubsection{Micro-instructions and Sub-instructions}
Child-ILAs can be used to model specifications or implemented algorithms. 
When modeling an implemented algorithm, their instructions serve the same role as {\bf micro-instructions} for complex instructions in processors, which represent one possible implementation of that instruction. 
We distinguish this from instructions of child-ILAs when used for specification, i.e., when they specify behavior that must hold for \emph{all} implementations. 
In the latter case, we call them {\bf sub-instructions}. 
For example, in \texttt{\small REP MOVS}, the steps of the instruction are part of the specification, and thus, these sub-instructions are required of {\it all} implementations. 
Therefore, child-ILAs will be referred to as sub- or micro-ILAs depending on whether they have sub- or micro-instructions respectively.
 
The distinction between micro-instructions (implemented algorithm) and sub-instructions (specifications) is important. 
For example, the ARM Cortex-M3 user guide~\cite{ARMCortexM3spec} says that load-multiple (\texttt{\small LDM}) and store-multiple instructions  (\texttt{\small STM}) access memory ``in order of increasing register numbers.'' 
Another shared memory interacting processor would expect to see this order.
Further, these instructions are interruptible, thus the intermediate values of the architectural states are visible to the interrupt handler.
This order of accesses is therefore desired in the formal abstraction when verifying systems with multiple interacting hardware components. 
However, while the user guide only describes one particular implementation, the ARM architecture specification does not impose this ordering requirement. 
This is reflected in the previous work on the ARM ISA formal specification and verification~\cite{arm16cav}. 
While they state that ``some load instructions may be split into multiple micro-ops'' and account for it by updating the architectural state when each micro-op completes, when verifying this instruction they check the state  only ``when the last micro-op completes.'' 
We emphasize that it is important to treat these split accesses as \emph{micro-instructions} (i.e., as implemented algorithms) and not \emph{sub-instructions}. 

Due to the differing roles of specifications and implementations in verification, we impose some restrictions on hierarchical ILAs. A sub-ILA may contain sub-ILAs or micro-ILAs. However, a micro-ILA can only contain micro-ILAs, as an implementation cannot contain a specification. 

\subsubsection{Definition of Hierarchical ILAs}

A hierarchical ILA $A$ is defined as: $ \langle S, I, W, V, F, D, N, C \rangle$. The new component $C = \{ (A^c_1, m_1), \dots \}$ is a set of tuples consisting of child-ILAs and a Boolean flag that denotes whether the particular child-ILA is a micro-ILA ($m_i = \texttt{true}$) or a sub-ILA ($m_i = \texttt{false}$).

In Figure~\ref{fig:aes-ila}, Part (b) shows the ILA definitions for $S$, $I$, $W$, $V$, $F$, and the decode ($\delta_i$) and state update functions ($N_i$) for two of the instructions in the AES example.
Part (a) of Figure~\ref{fig:aes-child-ila} shows the definitions for a child-ILA that models the \texttt{\small{START_ENCRYPT}} instruction.

\subsection{ILA Execution Semantics}
\label{sec-formal-exe}
An ILA model is essentially a labeled state transition system that emphasizes modularity through a set of instructions.
The semantics of execution of an ILA instruction is as follows: 
\begin{equation}
  \infer{S \overset{i}\rightsquigarrow S'}
        {V(S, W) & \delta_i(F(S, W)) & S' = N_i(S, W)}
  \label{rule-ILA-exec}
\end{equation}
Rule~(\ref{rule-ILA-exec}) says that an ILA can transition from state $S$ to $S'$ if the following conditions are satisfied:
\begin{itemize}
  \setlength\itemsep{0em}
  \item An instruction is triggered: $V(S, W)$ is $true$.
  \item The $i$-th instruction is issued: $\delta_i(F(S, W))$ is $true$.
  \item State update of the vector $S'$ is according to $N_i(S, W)$.
\end{itemize}

Execution of a child-instruction in a child-ILA is similar: 
\begin{equation}
  \infer{S^c \overset{j}\rightsquigarrow S'^c}
        {V^c(S^c, W^c) & \delta^c_j(F^c(S^c, W^c)) & S'^c = N_j^c(S^c, W^c)}
  \label{rule-cILA-exec}
\end{equation}
State updates in instructions at the lowest level of an ILA hierarchy are considered to be \emph{atomic}, i.e., indivisible. This enables reasoning about concurrency with multiple ILAs.

The focus of this paper is on using an ILA to model the behavior of a single processor/accelerator core using instructions. This is useful for capturing a \emph{sequential programming model} for the core's operation as it processes a \emph{sequence} of instructions. 
Although the hardware may operate on instructions in parallel (similar to pipelined processors), the programming abstraction for software is that of a single sequential thread of control (similar to the ISA programming model). The value of the ILA is that this sequential programming model is now extended uniformly from processors to hardware accelerators. \emph{We believe this abstraction from parallel hardware in accelerators to a single sequential programming model is a key enabler for system design and verification, and a central contribution of the ILA methodology.}

Further, once we have ILAs, each of which represents a single thread of control that updates shared architectural state, we can use them to model a system of concurrent cores with shared memory. 
Specifically, instructions are sequentially composed within an ILA, whereas concurrency and interleaving models are handled outside of ILAs. 
Analogous to ISAs for processors, we can use techniques for modeling multi-thread concurrency and memory consistency with multiple ILAs. 
This is discussed briefly in Section~\ref{sec:mcm} later -- case studies and applications with concurrent cores are outside the scope of this paper.

\section{Case Studies: Modeling} 
\label{sec:case}
In this section, we evaluate the ILA's modeling abilities using four case studies: application-specific accelerators for image processing, machine learning, and cryptography; and the Rocket processor core based on the RISC-V ISA. 
With designs from different application domains, the ILA is shown to be a uniform model usable across heterogeneous accelerators and processors. 
Verification for these case studies is described in the next section.

We create the ILA for each design based either on an informal English specification or a high-level reference model.
These ILAs are synthesized using template-driven program synthesis~\cite{subramanyan2017template}, or in some cases manually written in Python using our ILA library API. 
\footnote{All models and templates are available on GitHub:
\url{https://github.com/PrincetonUniversity/ILA-Modeling-Verification}.}
Table~\ref{tab:ilacase} provides information about each case study. Columns 2-5 give the reference model type, and sizes of the reference model and RTL, respectively. 
The RTL descriptions are either generated by high-level synthesis or taken from OpenCores.org. 
Columns 6 and 7 provide the number of instructions/child-instructions in the ILA, and ILA size (in lines of Python code). We now discuss salient aspects of each case study.
\begin{table*}[!htbp]
\footnotesize
\centering
  \begin{threeparttable}
\caption{ILA Modeling Case Studies}
\label{tab:ilacase}

\begin{tabular}{lllllcl}
\toprule
\multicolumn{1}{c}{\multirow{4}{*}{\bf Design Name}} & \multicolumn{4}{c}{\bf Design Statistics}                          & \multicolumn{2}{c}{\bf ILA}                                                                                                                                            \\ \cmidrule(lr){2-5} \cmidrule(lr){6-7} 
\multicolumn{1}{c}{}     & {\bf Reference}     & {\bf Ref. Lang.} & {\bf Ref. Size}      & {\bf RTL Size} & \multicolumn{1}{l}{\begin{tabular}[c]{@{}l@{}}{\bf \# of insts.}\\ {\bf (parent/child)}\end{tabular}} & \begin{tabular}[c]{@{}l@{}}{\bf ILA Size}\\ {\bf (Python LoC)}\end{tabular} \\ \midrule
RBM  			& System-level design~\cite{pilato2016design} & SystemC & 1211$^\dagger$ & 10578    & 3/14                                                                                            & 1009                                                              \\ 
GB (High-level) & Halide description~\cite{pu2016programming} & C++ & 288$^\dagger$  & 6935     & 2/2                                                                                             & 538                                                               \\ 
GB (Low-level) 	& HLS input~\cite{pu2016programming} & C++ & 1718$^\dagger$ & 6935     & 2/4                                                                                             & 1561                                                              \\ 
AES (table) 	& RTL simulator~\cite{ocAES} & C++ & 1905           & 1105     & 8/5                                                                                             & 435$^\star$                                                       \\ 
AES (logic) 	& Software simulator~\cite{ocAES}  & C             & 328            & -        & 8/7                                                                                             & 337$^\star$                                                       \\ \midrule
RISC-V Rocket            & Chisel description~\cite{asanovic2016rocket} & Chisel & 3488$\ddagger$ & 18252    & 43                                                                                              & 1672$^\star$                                                      \\  \bottomrule
\end{tabular}
    \begin{tablenotes}
    \item $\star$ILA synthesis template size. $\dagger$Excluding shared library. $\ddagger$Processor core only.
    
    \end{tablenotes}
\end{threeparttable}
\end{table*}

\subsection{Application-Specific Accelerators} \label{sec-case-acc}
We consider two types of accelerators: (i) those using local memory for computation and direct memory access (DMA) to load/store data into their local memory buffers, and (ii) those streaming input and output data. The commands at the interface 
relate to: (i) the interface protocol, and (ii) the computation tasks. 
In the AES example, the interface protocol refers to setting configurations and querying the status, and the computation task is the block encryption operation modeled in the \texttt{START_ENCRYPT} instruction.
\subsubsection{\bf Restricted Boltzmann Machine}
Restricted Boltzmann Machine (RBM) is a stochastic neural network commonly used in recommendation systems. We model the RBM accelerator from the Columbia System Level Design Group~\cite{pilato2016design}. It is implemented in SystemC and synthesized to Verilog. The accelerator supports both prediction and training, and uses the contrastive divergence learning algorithm. It exchanges data with shared memory via DMA. 

We manually constructed the ILA of the RBM accelerator. The ILA captures both the interface protocol and the computation. It models the interface activities where the accelerator autonomously initiates DMA transactions to load and store training/testing datasets after receiving an initial configuration. It contains 3 instructions, \textit{ConfDone}, \textit{ReadGrant}, and \textit{WriteGrant}, which set the configuration, grant DMA read and write transactions, respectively. The complexity of computation and DMA interaction is managed by five child-ILAs for loading, storing, coordination, training, and prediction, respectively, comprising a total of 14 child instructions. The training and predicting child-ILAs, in turn, have child-ILAs that model their computation. The computation iteratively updates two regions of private local memory for the hidden layer and visible layer in a fixed order. This order is maintained by control registers in the implementation, using child-ILA states. 
Recall that child-ILA states are updated by a child instruction, which activates the decode function of a subsequent child instruction.

This case study illustrates handling of both protocol and computation, the value of hierarchical ILAs, and how order is captured by the state update and decode functions of the child-ILA.

\subsubsection{\bf Gaussian Blur}
The image processing accelerator performing the Gaussian Blur (GB) operation is from the Stanford VLSI Research Group~\cite{pu2016programming}. Its behavior is described in Halide~\cite{ragan2013halide}, a domain specific language for developing high-performance image processing applications. Halide descriptions can be compiled into C++, which can then be synthesized to a Verilog implementation through high-level synthesis (HLS). The GB accelerator takes an image as streaming input, and utilizes a 2-dimensional \textit{line buffer} to collect one part of the image at a time for the GB kernel function computation. It then streams out the result for each part as soon as it is ready. 

We manually construct two ILAs, GB$_H$ and GB$_L$, from design descriptions at two different levels. GB$_H$ is derived from the high-level Halide description, and models the specification. GB$_L$ is derived from the lower-level C++ code compiled from the Halide description and models micro-architectural details. 
GB$_H$ captures the size of input and output images, the streaming pattern (row-major traversal), data \textit{source} for the kernel function, and \emph{when} the result is ready. GB$_H$ does not specify how streamed data is buffered, whereas GB$_L$ additionally includes a specific \textit{line buffering} mechanism~\cite{pu2016programming}.  

In this case study, we focus on specifying the streaming data interface and the output image accumulation. The kernel computation is modeled as an uninterpreted function, a standard practice in verification to allow  decoupling of control verification from data-intensive computations that can be verified separately. (This is supported by standard SMT solvers, described in Section~\ref{sec:smt}.) Both GB$_H$ and GB$_L$ have two instructions, \textit{WRITE} and \textit{READ}, that represent sending and receiving a pixel to and from the I/O boundary, respectively. The two ILAs have the same instruction set, i.e., the same hardware interface, but have different levels of abstraction. The extra complexity of GB$_L$ in modeling the 2-dimensional line buffer and stream buffers is captured by its child-ILAs; child-instructions model data movement between different components. 

This case study serves to illustrate the ability of the ILA to model: (i) streaming I/O, and (ii) different levels of abstraction for the same instruction set through additional micro-architectural detail.

\subsubsection{\bf Advanced Encryption Standard}
This case study, introduced in Section~\ref{sec-formal-def}, considers a cryptographic engine from OpenCores~\cite{ocAES} implementing the Advanced Encryption Standard (AES). 
The accelerator receives configurations via memory-mapped I/O and uses DMA to exchange data with shared memory. 
The configuration includes the encryption key, initial counter value, plaintext location, and length, which are stored in registers mapped to the memory address space. 
This accelerator works in  AES-CTR mode~\cite{lipmaa2000comments}, where the plaintext is fetched from the shared memory starting from the location pointed to by the \textit{plaintext location}. 
The accelerator operates in the following sequence: fetch one block from memory, apply exclusive-OR operation between plaintext and the AES encrypted counter to get the ciphertext, and then store the block back into the same location. 
Each block has 128 bits and the complete encryption operation has 10 rounds. 
The ILA model uses child-ILAs for modeling the encryption function.

We synthesize two different ILAs using template-driven synthesis~\cite{subramanyan2017template}. These ILAs, AES$_C$ and AES$_V$, are based on C and Verilog implementations, respectively. They have the same architectural instruction set, 
but with differences in the block-level and round-level implementations in their micro-instructions.

The instructions on the interface has been shown in Figure~\ref{fig:aes-ila}. Only \texttt{START_ENCRYPT} instruction has child instructions, depicting  block-level encryption. 
At the block level, AES$_V$ has more child state variables (mostly counters and control signals), and its memory access is modeled at a finer granularity than AES$_C$. 
At the round level, there is one micro-instruction for each round.
AES$_V$ uses a table look-up, while AES$_C$ uses logical operations. We capture these differences in 
their micro-ILAs. This case study illustrates the ability of the ILA to describe two different implemented algorithms for the same set of instructions. 



\subsection{General Purpose Processors}
The ILA of a general purpose processor is based on its ISA, and the ILA has the same instructions and semantics. However, in contrast to existing formal ISA models (e.g., ISA-Formal~\cite{arm16cav}), 
our model has a uniform treatment of interrupts (and possibly other input signals) and instructions, rather than treating interrupts as a special case.
Further, it supports hierarchy and distinguishes sub-instructions from micro-instructions; this is missing in previous work.

\subsubsection{RISC-V} 
RISC-V is a free and open ISA with increasing adoption in industry and academic research. 
It has a base ISA with several extensions for advanced functionality. 
We synthesize the ILA of the base integer ISA RV32I with the \texttt{\small DefaultRV32Config} of  Rocket -- a single-issue in-order 5-stage pipeline implementation (part of the Rocket Chip SoC generator)~\cite{asanovic2016rocket}.

The ILA covers: (1) user-level base integer registers and instructions, (2) machine-level control status registers (CSRs), (3) environment call/trap return instructions, (4) the address translation and the memory-management fence, and (5) interrupt and hardware interrupt handling. 
The semantics of each instruction are as follows: if an interrupt occurs, the next state is updated as the result of the interrupt. Otherwise, the state update is performed according to the instruction word. 
This case study demonstrates modeling interrupts and instructions uniformly. 
The RISC-V ISA exposes the synchronization between the memory hierarchy and the translation lookahead buffer (TLB) through the \texttt{SFENCE.VMA} instruction.
The lack of synchronization could result in stale page table entry (PTE) references. 
The TLB in the RISC-V ISA is software visible, and we include it in our ILA model as an architectural state variable. 
However, its size, associativity and other parameters are not specified by the ISA specification, so we model it as a ghost TLB,
which can potentially hold any PTE that has been referred to but has not been explicitly flushed out. 
As a 32-bit RISC-V model, it only models the \texttt{Sv32} virtual addressing in addition to \texttt{Bare} mode.
Memory consistency issues are beyond the scope of the current case study and thus not modeled. (Memory consistency is briefly discussed in Section~\ref{sec:mcm}.)


\vspace*{-.05in}
\subsection{Summary}
From these case studies and the data in  Table~\ref{tab:ilacase}, we make the following observations:
\begin{itemize}
    \setlength\itemsep{0em}
    \item Accelerator ILAs tend to have a small number of instructions/child-instructions. That is, most accelerators can be specified by just a handful of instructions. 
    \item The same design can be modeled using ILAs at differing levels of detail. 
    (In the next section we show how these different models are checked for equivalence.)
    \item The ILA model (or template, when the ILA was synthesized) has size comparable to a reference design in C/SystemC/C++/Chisel. Thus, the value of its \emph{formal} model comes at no additional cost, in terms of the size of a reference description.
    \item The ILA model (or template) is \emph{significantly smaller} than the final RTL implementation, making this an attractive 
    entry point for verification and validation.
\end{itemize}
\section{Case Studies: Verification\label{sec-4}}

\begin{table*}[!htbp]
\footnotesize
\centering
  \begin{threeparttable}
\caption{ILA Verification Experiments}
\label{tab:ivcase}
\begin{tabular}{llllll}
\toprule
{\bf Category}               & {\bf Designs}                      & {\bf Models}                        & {\bf Tools}                 & {\bf Strength of Proof}           & {\bf Time} \\ \midrule
\multirow{2}{*}{ILA vs. ILA} & \begin{tabular}{l}GB\end{tabular} 								& GB$_H$ vs. GB$_L$                   & JasperGold                  & complete                          & 2h 27m \\  
                             & \begin{tabular}{l}AES\end{tabular}					 			& AES$_C$ vs. AES$_V$                 & ILA lib+JasperGold$^\dagger$& complete                          & 15m  \\ 
                            \midrule
\multirow{8}{*}{ILA vs. FSM} & \multirow{2}{*}{\begin{tabular}{l}GB\end{tabular}}                & GB$_H$ vs. Verilog                  & JasperGold                  & complete                          & 2h 50m  \\ 
                             &                                    & GB$_L$ vs. Verilog                  & JasperGold                  & complete                          & 16h 12m \\ \cmidrule(lr){2-6}
                             & \multirow{2}{*}{\begin{tabular}{l}RBM\end{tabular}}               & ILA vs. SystemC                     & CBMC                        & complete                          & 2h 7m\\ 
                             &                                    & ILA vs. Verilog                     & JasperGold                  & complete                          & 6h 54m \\ \cmidrule(lr){2-6} 
                             & \multirow{3}{*}{\begin{tabular}{l}RISC-V\\Rocket\end{tabular}}     & \multirow{3}{*}{ILA vs. Verilog}    & \multirow{3}{*}{JasperGold} & complete (invariants)                 & 5h 40m \\  
                             &                                    &                                     &                             & complete (interrupt)              & 8m  \\  
                             &                                    &                                     &                             & BMC to 40 cycles (instructions)   & 86h 5m \\ 
                             \bottomrule

\end{tabular}
    \begin{tablenotes}
\item 
$^\dagger$ILA library (using Z3) for block-level ILA equivalence, JasperGold for round-level equivalence. 
    \end{tablenotes}
\end{threeparttable}
\end{table*}
The ILA model can represent specifications or implementations of hardware modules. 
In this paper, we focus on using ILAs for 
hardware verification to check that implementations of accelerators/processors match their ILA architectural specifications. This also enables checking that different implementations of an accelerator have the same behavior at their interface specified by an ILA, thereby proving their architecture-level equivalence. 

We briefly touch on the underlying formal verification techniques, then discuss ILA-based verification, and finally describe their evaluation on our case studies.

\subsection{Underlying Formal Verification Techniques}
\label{sec:smt}
SMT solvers~\cite{smt-handbook-09, smt-demoura-2011} provide decision procedures for first-order logic formulas in background theories, and have found numerous applications in verification. 
In this work, we use quantifier-free formulas that use the theories of arrays, uninterpreted functions and bitvectors (\texttt{QF\_AUFBV} in the SMTLIB standard~\cite{smtlib-v2-10}). 

Model checking is a verification technique to check correctness properties for a finite state transition system~\cite{cgp-99, smc-93}. 
Unbounded model checking explores all reachable states of the transition system while bounded model checking (BMC)~\cite{bmc-03} restricts the search to all states reachable within the first $k$ transitions of the system. $k$ is referred to as the \emph{bound} and is typically set by the verification engineer. BMC alone cannot prove the absence of property violations, however it is very effective for bug finding in practice~\cite{copty-cav-01}.\footnote{The success of BMC is often ascribed to the ``small world hypothesis'': bugs (inadvertent mistakes, as opposed to maliciously introduced design flaws) are likely reachable through \emph{some} short sequence of steps from the initial state.} 

\subsection{ILA-based Verification \label{sub-sec-model-eq}}


As described in Section~\ref{sec-formal-def}, 
the ILA model is a labeled state transition system, but one that emphasizes modularity and hierarchy. 
These features simplify verification through decomposition along (child-)instructions and architectural state elements. 
We consider two main settings for ILA-based verification:  
(i) ILA vs. ILA, and (ii) ILA vs. FSM. The equivalence of these models is based on bisimulation relations on the underlying labeled state transition systems~\cite{Milner89}. (It is also straight-forward to consider stuttering in addition, or extend our discussion to model refinement by using simulation relations and containment checks instead.) 

\subsubsection{ILA vs. ILA verification}
As the GB  and AES case studies described in Section~\ref{sec-case-acc} illustrate, we can construct ILAs for designs with differing implementations, or even at different levels of abstraction. A natural application is to check these ILAs for equivalence. In this setting, we compare two ILAs with the same instructions and sub-instructions, but with possibly different micro-instructions in the implementation.
For ILAs, instruction-based modularity provides the basis for establishing correspondence between two models, i.e., we check that the behavior of the ILAs is the same for each instruction and sub-instruction. 



Consider first the case where we do not have micro-instructions (implementations) in the ILA models.
Given ILAs $X$ and $Y$, we check that the issuing condition and the next-state transition updates for each instruction and sub-instruction are equivalent in the two models. 
Specifically, the equivalence for (sub-)instruction $i$ is verified by checking:
\begin{enumerate}
\item[(i)] equivalence of the valid function: $\forall S, W.~(V^X(S, W) \leftrightarrow V^Y(S, W))$
\item[(ii)] equivalence of the decode function: $\forall S, W. ~(\delta^X_i(S, W) \leftrightarrow \delta^Y_i(S, W))$
\item [(iii)] equivalence of state updates: $\forall S, W. ~(\delta^X_i(S, W) \wedge \delta^Y_i(S, W) \to (N^X_i(S, W) = N^Y_i(S, W)))$.
\end{enumerate}
Note, $X$ and $Y$ are shown with the same state variables here, but this can be generalized to a mapping between their variables.

Now consider the case where we have micro-instructions in the ILA model(s), to represent micro-architectural implementation choices. We do not enforce equivalence at the micro-instruction level. Instead we check the equivalence of each instruction and sub-instruction, where each may be implemented using a sequence of micro-instructions.
Here, we check equivalence \emph{after} the sequence of micro-instructions that implements a instruction/sub-instruction is completed. 


To check the equivalence for each instruction, we may need additional abstraction/refinement mappings to establish "corresponding" states between the two models. Thus, the equivalence check essentially says that if we start in corresponding states and apply an instruction, then we end in corresponding states. Here, we leverage well-studied processor verification techniques~\cite{burch1994automatic, jhala2001microarchitecture} that propose and use such mappings. 
In addition, we use invariants to prune some unreachable micro-architectural states, such as the invalid combination of the horizontal/vertical frame pointers of an image in the GB case study.
These are often needed to prove the correspondence checks. 

\subsubsection{ILA vs. FSM Verification \label{sec-ila-vs-fsm}}
In this setting, we are interested in verifying that a hardware implementation available as an FSM model (e.g., RTL) corresponds to its ILA specification. 
As before, the equivalence between an ILA model and an FSM model is checked for each instruction and sub-instruction in the ILA. However, unlike the ILA that has a clear set of instructions, an FSM model is generally  a monolithic transition system without a separation between the parts implementing different instructions/sub-instructions. 

Again, we leverage well-studied processor verification techniques~\cite{burch1994automatic, jhala2001microarchitecture}, and use refinement mappings to relate the FSM states to the ILA states for each (sub-)instruction. 
Invariants are also used to prune unreachable micro-architectural states in the FSM model, e.g., the invariant on an one-hot encoded counter in the RTL implementation. 

\emph{Note that ILAs enable a discipline for accelerator implementation verification that is based on established methodology for processor verification.} This is in contrast to customizing general hardware verification techniques for this task, since determining what/when to check is itself a challenge and in practice woefully incomplete.

\subsection{Experimental Evaluation}
We have implemented the ILA-based verification techniques described above, on top of off-the-shelf verification tools (Z3~\cite{de2008z3}, JasperGold~\cite{jasper-url} and CBMC~\cite{clarke2004tool}).
The correspondence checks on instructions are expressed as verifying the assertions where the two models end in corresponding states, given the assumptions that they start in corresponding states and apply the same instruction. 
For example, for the Gaussian-Blur accelerator, we check the correspondence of frame pointers by assuming the two models initially have an equal horizontal/vertical frame pointer pair.
Then, we verify that their frame pointers are equal after the \textit{WRITE} instruction, regardless of the pixel accumulating and buffering mechanism. 
Our ILA library supports translation of the ILA models into formats supported by these tools. Verification results are summarized in Table~\ref{tab:ivcase}.


\subsubsection{ILA vs. ILA Verification} ~\\
{\bf \emph{GB Accelerator}}: Recall that we constructed two ILAs (see Section~\ref{sec-case-acc}) for the GB accelerator. The ILA GB$_H$ follows the high-level Halide code, and GB$_L$ follows the lower-level C++ code. The two ILAs have the same instruction set, but GB$_L$ has additional micro-instructions to describe stream buffer and pixel accumulation operations. 
We check equivalence of each instruction using the Burch-Dill approach~\cite{burch1994automatic}, 
where we use a ``flushing" function to relate corresponding states in the two ILAs. This is needed to abstract away intermediate micro-architectural states in GB$_L$ that are not visible in GB$_H$. Specifically, the checked instruction starts in a state in GB$_L$ where there is no buffered intermediate data. Thus, for each instruction, we check that the architectural states (IO ports, image frame, pixel pointers, etc.) are equal at the end, whenever the ILAs start in corresponding initial states.
Verification completed in about 2.5 hours using JasperGold.

\vspace{2pt}
\noindent
{\bf \emph{AES Accelerators}}: 
The two ILAs of the AES accelerator were described in Section \ref{sec-case-acc}. They have the same instructions at the top level, but different micro-instructions due to different implementations of the encryption algorithm.
We leveraged the hierarchy in ILA models 
to decompose equivalence checking into block-level and round-level equivalence checks. 

The AES encryption is a ten-round operation.
As both models have one micro-instruction for each round, we first check the equivalence of such micro-instructions. 
At the round level, we check that the generated round keys and ciphertexts are matched after the execution, given their round keys and cipherstates are matched before the execution. 
The micro-instructions and the verification conditions for checking are automatically converted into Verilog to take advantage of hardware verification tools, which are better at reasoning logic operations in AES encryption. 
Based on the equivalence of round-level micro-instructions, we check the equivalence of the ten-round AES operation by modeling the round-level encryption as an uninterpreted function. 

The block-level operations involve fetching plaintext, encrypting data, storing ciphertext, and maintaining encryption states, e.g., the counter. 
We check that after processing one block, the two models should have the same ending state (including shared memory and registers in the accelerator) if they start from the same state. 
By proving the equivalence of the micro-instructions performing block-level operations, the equivalence of \texttt{START_ENCRYPT} instruction, which processes series of blocks, can be guaranteed. 
We used our ILA library, which in turn uses the Z3 SMT solver~\cite{de2008z3}, for checking block-level equivalence, and use JasperGold for checking round-level equivalence. 
The total verification time was about 15 minutes. 

These two case studies show that ILA equivalence checking can be applied to bridge the gap between models at different abstraction levels associated with design languages (Halide vs. C++, C vs. Verilog). 

\subsubsection{ILA vs. FSM  Verification}
We consider FSM models at the register transfer level (e.g., in Verilog) or system level (e.g., in C/SystemC).
We check ILA vs. FSM equivalence for two accelerators and a general purpose processor.
All FSM models are provided independently by other groups, and not synthesized from ILAs: RBM-SystemC model by the Carloni-Columbia group~\cite{pilato2016design}; Gaussian-Blur-Halide/C++ model by the Horowitz-Stanford group~\cite{pu2016programming}; AES-C/RTL implementation from OpenCores.org~\cite{ocAES}; RISC-V implementation from Berkeley's Rocket-chip generator~\cite{asanovic2016rocket}. 
Our previous work~\cite{subramanyan2015template,subramanyan2017template} has discussed the verification of the 8051 micro-controller and SHA accelerator, where 8 bugs were found in the RTL model in 8051.

\vspace{2pt}
\noindent
{\bf \emph{GB Accelerator}}:
We performed equivalence checking between the RTL implementation (generated by HLS) and each of the two ILAs, GB$_H$ and GB$_L$, separately. 
GB$_L$ models more detailed behavior such as buffering and pixel accumulation, which is similar to the RTL implementation.
We provided invariants to establish corresponding states, and successfully completed verification against each ILA model. As expected, the verification of RTL against the more detailed GB$_L$ took more time than against GB$_H$ ($\approx 16$ hrs vs. $3$ hrs).

\vspace{2pt}
\noindent
{\bf \emph{Restricted Boltzmann Machine}}:
We exploited the structural similarity between SystemC, Verilog, and the ILA models to expedite equivalence checking through modular checking. 
We replaced some functions in the computation, e.g., the \emph{sigmoid} function, with uninterpreted functions. 
(Verification of these functions can be addressed separately.) 
We successfully completed verification of the ILA vs. SystemC ($\approx 2$ hrs), as well as the ILA vs. Verilog ($\approx 7$ hrs).
This example demonstrates that a single ILA can be matched against multiple FSM models 
with implementation-specific differences.

\vspace{2pt}
\noindent
{\bf \emph{ILA for RV32I vs. Rocket}}: 
We synthesized the ILA for the RISC-V specification and verified this against Verilog of the Rocket processor core generated from a Chisel description~\cite{asanovic2016rocket}.  
The verification settings can be found in directory \texttt{RISC-V/ILAVerif} in our GitHub repository.
Our focus was on the processor core, and we separate it from the memory system and the branch predictor. 
We abstract the branch predictor by constraining the interface of the processor core where any valid prediction can arrive in any cycle.

Our verification had three main steps.

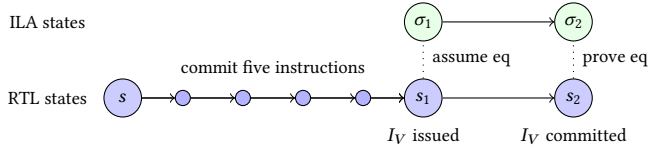
\begin{figure}[htbp]
  \begin{center}
    \begin{tikzpicture}
      \tikzstyle{rtl state}=[circle, fill=blue!20, draw, minimum width=0.5cm, inner sep=0pt]
      \tikzstyle{rtl ministate}=[circle, fill=blue!30, draw, minimum width=0.2cm, inner sep=0pt]
      \tikzstyle{ila state}=[circle, fill=green!10, draw, minimum width=0.5cm, inner sep=0pt]
      \newcommand{\pictext}[1]{\scriptsize{#1}}
      \node at (-1,1) {\pictext{ILA states}};
      \node at (-1,0) {\pictext{RTL states}};
      \node[ila state] (i1) at (4, 1) {\pictext{$\sigma_1$}};
      \node[ila state] (i2) at (6, 1) {\pictext{$\sigma_2$}};
      \node[rtl state] (s) at (0,0) {\pictext{$s$}};
      \node[rtl state] (s1) at (4,0) {\pictext{$s_1$}};
      \node[rtl state] (s2) at (6, 0) {\pictext{$s_2$}};
      \draw[->] (s) -- node[above, yshift=0.2cm] {\pictext{commit five instructions}} (s1);
      \node[rtl ministate] (sa) at (0.8, 0) {};
      \node[rtl ministate] (sb) at (1.6, 0) {};
      \node[rtl ministate] (sc) at (2.4, 0) {};
      \node[rtl ministate] (sd) at (3.2, 0) {};
      \draw[->] (s) -- (sa);
      \draw[->] (sa) -- (sb);
      \draw[->] (sb) -- (sc);
      \draw[->] (sc) -- (sd);
      \draw[->] (sd) -- (s1);
      \draw[->] (s1) -- (s2);
      \draw[->] (i1) -- (i2);
      \draw[dotted] (s1) -- node[right] {\pictext{assume eq}} (i1);
      \draw[dotted] (s2) -- node[right] {\pictext{prove eq}} (i2);
      \node at ($(s1.south) + (0, -0.25)$) {\pictext{$I_V$ issued}};
      \node at ($(s2.south) + (0, -0.25)$) {\pictext{$I_V$ committed}};
    \end{tikzpicture}
  \end{center}
  \caption{ILA vs. FSM Verification of Instruction Execution in Rocket.}
  \label{fig:riscv-verif}
\end{figure}

\begin{enumerate}
  \item First, as discussed in Section~\ref{sec-ila-vs-fsm}, we use implementation invariants to prune unreachable states in per-instruction equivalence checking. 
The first category of invariants targets the correctness of the bypassing network. 
That is, for each general purpose register, the value bypassed to the decode stage must be the same as the corresponding when the instruction at that stage commits.
The second category ensures the multiplication/division unit and co-processors do not generate valid response signals when executing integer instructions. 
These invariants are verified using the unbounded model checking engines of JasperGold.
  \item Next, we verified interrupt handling. The RTL handles interrupts by inserting dummy instructions in the pipeline, corresponding to the interrupt instruction in the ILA. We proved that RTL and ILA states match when the interrupt commits  (using JasperGold).
  \item Finally, we checked equivalence on ordinary instructions using the inductive proof strategy shown in Figure~\ref{fig:riscv-verif}. The processor is started in an arbitrary state $s$ constrained by the invariants described in Step (1). Five\footnote{In our experiments, five instructions led to an over-approximation of the reachable states that is ``strong'' enough to prove equivalence. }
  instructions are issued, leading to a state $s_1$ where we assume that they have been correctly committed, i.e., the ILA state $\sigma_1$ and RTL state $s_1$ are equal. Then, a new instruction $I_{V}$ is issued, and we check whether ILA state $\sigma_2$ and RTL state $s_2$ match when $I_{V}$ commits. We were unable to complete an unbounded proof of this property. However, except for the bug discussed below, there was no violation up to a bound of 40 cycles using BMC (from $s$ to $s_2$). 

Note that the latency of an instruction depends on the response latency from modules like the data cache. Therefore, it is possible that 40 cycles are not sufficient to \emph{guarantee} that $I_V$ commits correctly. Future work will build a memory model that can prove full correctness to avoid this limitation.
 
\end{enumerate}

The two main challenges in verifying the Rocket core are finding a sufficiently strong set of invariants so that the inductive proof (step 3) above succeeds and specifying the refinement relation. 

\noindent \textbf{\textit{Deriving Pipeline Invariants for Rocket Verification:}}
We derived the ``strengthening'' invariants using a counter-example guided approach. 
Initially, we attempted the inductive check for instruction equivalence starting with an unconstrained state (i.e., no invariants). This resulted in spurious counter-examples where the inductive proof failed when starting from unreachable states. Analysis of these states helped us formulate the set of invariants described in Step (1). These invariants were checked using the unbounded model checking and then used to constrain the starting state for the inductive proof described in Step (3) above.
The base case of the inductive proof: ensuring that five instructions commit correctly starting from the reset state was verified separately.

\noindent \textbf{\textit{Deriving Refinement Relations for Rocket Verification:}}
As discussed in Section~\ref{sub-sec-model-eq},
the equivalence in Figure~\ref{fig:riscv-verif} is defined with respect to refinement,  which consists of a mapping between the states in Rocket implementation and our ILA model. The states here involve general purpose registers, control and status registers (CSRs), the program counter and  memory. The refinement relations for each of these state variables are as follows:
\begin{enumerate}
	\item General purpose registers and CSRs: 
This refinement relation specifies that the register values in the ILA model and the Rocket implementation must be equal after each instruction commits.
    \item Program counter: 
The program counter's refinement relation is a little trickier due to branch prediction and speculative execution.
In the Rocket implementation, every pipeline stage except the fetch stage possesses a program counter variable corresponding to the current instruction in that stage. 
The refinement relation for the PC specifies that the program counter value of the commit stage of the Rocket implementation and the corresponding program counter value in the ILA model should be equal after each instruction commits. 
    \item Memory: 
Instead of modeling two individual memories and checking value equivalence over all addresses, we use a shared memory for all memory read operations, and store all the memory write operations separately for comparison. 
Equivalence requires that the changes to memory should be the same when $I_{V}$ commits. 
We abstract the memory for read operations by returning arbitrary values for irrelevant read requests, and only enforcing the equivalence on the requests from $I_{V}$. 
\end{enumerate}


To track the stages where current instruction $I_{V}$ resides, we use a sequence monitor to store the corresponding stages and use these in specifying the refinement relations. 

\begin{figure}
\centerline{\includegraphics[width=0.3\textwidth]{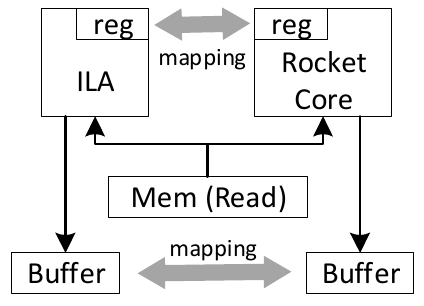}}
\caption{ \label{fig:rv-compound} The compound transition system used in ILA vs. Rocket processor core equivalence checking. }
\end{figure}

The compound transition system in Figure~\ref{fig:rv-compound} shows how we show that the Rocket implementation refines the ILA model.
This compound transition system has more than 4k bits of flip-flops and 320k gates (reported by JasperGold).  JasperGold uses both bounded and ubounded model checking techniques (both bounded and unbounded) on this transition system and any trace violating the refinement relations indicates the two models are not equivalent. 
\noindent \textbf{Rocket Implementation Bug:}
We found a bug where the Rocket core incorrectly implements the trap return instructions. According to the specification~\cite{RISCVSpec} these instructions should set the \textit{xPIE} bits in \textit{mstatus} register to 1. However, the implementation sets them to 0. We reported this bug, and it has been fixed since. 
This case study illustrates usefulness of our approach on real processors. 


\subsection{Summary of Verification Experiments}
From Table~\ref{tab:ivcase}, we observe that
most verification experiments can derive a complete proof where ``complete'' refers to either unbounded proofs, or running BMC to the upper bounds of the instructions' latencies.
Instructions on the Rocket core are checked up to a bound of 40 cycles, which is incomplete, but does provide a significant level of assurance. 


Overall, our evaluation confirms the viability of equivalence checking using ILAs, where we leverage the ILA modularity and hierarchy on top of existing verification tools and processor verification methodology, to successfully verify a range of accelerators and processors. 
Our case studies cover the verification of computation (AES and RBM) as well as processor/accelerator interfaces (GM and RBM), which is important for accelerator verification. 

\section{Discussion and Future Work}
\label{sec:applications}

While this paper focuses on the application of the ILA model in verification of a single compute engine (processor or accelerator), the ILA has other applications as discussed below.

\subsection{Modeling Concurrency and Memory Consistency}
\label{sec:mcm}


The ILA model views compute engines (processors and accelerators) as processing a sequence of instructions. Although the underlying hardware may operate on these instructions in parallel (similar to pipelined microarchitectures for processors), the programming abstraction it provides is that of a single sequential thread of control (similar to the ISA-based programming model). 

As a next step, we believe that individual ILA models can be composed to perform reasoning over a \emph{concurrent} system of multiple accelerator/processor cores. Here, the large body of work on concurrent programs and multiprocessor systems can be leveraged, and potentially extended to accelerator-rich systems using ILAs. One natural application is to use concurrent program verification techniques for checking correctness properties at the system level. This would include use of well-known methods and tools, such as software model checking with partial order reduction~\cite{cgp-99, holzmannSpin} and compositional frameworks for thread-modular  reasoning~\cite{owicki1976axiomatic, threaderCav11}. 

Another promising application is in verification of memory consistency models, which capture rules about operations on shared memory. The ISA plays a central role in many efforts related to verification of memory consistency -- correctness of compiler mappings for higher-level languages~\cite{battyPopl11, battyPopl12}, correctness of microarchitecture implementations (including coherence and virtual memory subsystems) with respect to ISA and microarchitecture specifications~\cite{lustig2014pipecheck,lustig2016coatcheck}, and more recently at the trisection of software, hardware, and ISA~\cite{trippel2017tricheck}. Furthermore, there have been recent advances in automatic methods for verification~\cite{alglave2014herding} and synthesis~\cite{bornholt2017synthesizing} of axiomatic memory models. 

Note that these techniques and tools are not currently directly applicable to accelerators, where the hardware is described with low-level FSM models (e.g., RTL Verilog). More importantly, since the accelerator memory operations are generally not visible to the processor, ignoring these interactions with shared memory can have adverse consequences for checking correctness or security of the overall system. Modeling accelerator behavior as an ILA allows application (and potential extensions) of these known ISA-based techniques. We are currently working on memory consistency modeling for a general shared memory system with multiple processors and accelerators. 

Admittedly, this approach does not \emph{yet} address the challenging issues that currently pervade memory consistency verification using ISAs. However, it allows some separation of concerns, whereby good solutions for ISAs can be adapted for ILAs to extend their reach to accelerators. 

\subsection{Accelerator Code Generation}
Accelerators provide efficient hardware implementations of functions that can be offloaded from programmable processors. When accelerators are deployed, an important and error-prone task is to program the accelerator to invoke these functions. As discussed in Section~\ref{sec:intro}, the processor-accelerator interactions often use memory-mapped IO (MMIO). Even when a single ILA instruction implements a significant function (e.g., block encryption in the AES accelerator example), other instructions must precede this encryption instruction to set up the encryption key, address of the block, size of the block, etc. Thus, a sequence of instructions is needed to completely implement this function. This sequence is often referred to as the accelerator driver code and is typically provided as library code with the accelerator. For this code to be correct -- the instructions that set up the accelerator must be correct, as must the main accelerator function itself.
The ILA model enables correct code synthesis using well-known program synthesis techniques (e.g.,~\cite{Sygus13,jha2010oracle}). In this set-up, program synthesis seeks a program with $k$ ILA instructions that is equivalent to the software function $f$ it is replacing. Function $f$ serves as an oracle to guide the search, and the ILA model provides the accelerator instruction semantics for use in the SMT solver based search for the program with $k$-instructions. While this has not yet been implemented, the fact that these driver programs are short (i.e., $k$ is small) suggests promise for this useful application of the ILA model.

\subsection{Reliable Simulator Generation}
Given an ILA model, a reliable hardware simulator can be automatically generated for use in system/software development. 
The ILA model specifies the state update functions of the architectural state variables. Through hierarchy, it may optionally provide additional micro-architectural detail. 
These functions can be used to construct an executable model (i.e. a simulator) in almost any programming language (we currently use C++). 
As this simulator is generated from a formal specification that can be verified against the detailed RTL hardware model, this makes it a {\it reliable} executable model. 

Mismatches between a simulator and the hardware it models is a common problem for software (especially OS) developers. This problem can be addressed through generation of reliable simulators. As an illustrative example, we note that a previous version of the seL4 RISC-V port
makes no use of the supervisor memory-management fence (\texttt{SFENCE.VMA}) instruction,
but still executes  correctly on the spike ISA simulator. The simulator flushes the translation look-ahead buffer more frequently than either the Rocket implementation or the RISC-V specification's minimum requirement. 

We checked if the missing fence instruction would cause a problem. We removed the gratuitous TLB flushes in the simulator and embedded an address translation monitor to check whether any address translation uses a stale page table entry. The OS crashed on this modified simulator, and stale page table references were observed. This illustrates that the missing \texttt{SFENCE.VMA} could crash on a seL4  RISC-V port with a hardware implementation that conforms only to the minimum requirement in the specification. This mismatched behavior between the simulator and the hardware would be a problem if the OS were later ported to run on real hardware. Although the missing fence instruction has been added by the seL4 developers in a newer release, the simulator behavior of gratuitous TLB flushes has not been changed. The RISC-V community knows that the spike ISA simulator represents only one possible implementation of RISC-V, and that this might be different from a hardware implementation. However, we believe that it is useful to have an ISA-level simulator that represents the specification or matches a specific hardware implementation, so that software developers can be more confident about test results with the simulator.

\section{Related Work}
\label{sec:related}

To the best of our knowledge, this is the first work to formally model accelerator interfaces using the notion of instructions similar to ISAs for processors. Our previous work in~\cite{subramanyan2015template,subramanyan2017template} did introduce the notion of instructions as accelerator abstractions, but did not provide a formal model of execution. Instead its focus was on template-based synthesis of these abstractions. Further, these abstractions were defined as finite state transition systems, with no notion of hierarchy and no applicability to processors. In this work, we introduce the formal ILA model, with hierarchy (sub- and micro-instructions), that can be used uniformly across processors and accelerators. In addition, we provide an extensive evaluation of its modeling and verification capabilities on a diverse set of accelerator and processor designs. 
Past work~\cite{subramanyan2016verifying} has also shown how abstractions can be used for hardware/software co-verification. In contrast, the focus of this paper is on verification of hardware implementations against ILA specifications. 

Formal machine-readable and precise specifications~\cite{DBLP:conf/fmcad/Reid16, DBLP:conf/fmcad/GoelHKG14} of ARM and x86 processors have been developed. ISA-Formal~\cite{arm16cav} is a framework aimed at verification of ARM processors against ISA specifications~\cite{DBLP:conf/fmcad/Reid16}. However, as discussed earlier, this does not distinguish between different forms of hierarchy (sub-instructions vs. micro-instructions) needed for correct verification. Further, interrupts require special handling in their instruction semantics. Others~\cite{burch1994automatic, jhala2001microarchitecture} have targeted  verification of processor microarchitecture w.r.t. the ISA. These works target general purpose processors and do not address verification of accelerators. As discussed, we build on these techniques for verifying accelerator implementations against their ILA specifications. 


As discussed in Section~\ref{sec:background}, many efforts over the years have proposed the use of high-level models in design and verification. These include StateCharts, SystemC, Esterel, Transaction Level Modeling (TLM), BlueSpec, and
others~\cite{DBLP:journals/tosem/HarelN96,panda-systemc-01,DBLP:conf/iccad/BerryKS03, DBLP:conf/dac/BacchiniGMKDMGN07,hoe2000synthesis}. 
In particular, BlueSpec has been used as a high-level specification and design language in industry and research~\cite{shen1999using,nikhil-bluespec-04}.
It models hardware components as atomic rules of state transition systems and enables easy exploration of microarchitectural design space, e.g., adding a buffer in a pipeline.
The commercial BlueSpec compiler synthesizes the circuit implementation, i.e., Verilog, and exploits parallelism with a scheduler determining how to interleave the atomic rules~\cite{hoe2000synthesis,dave2007scheduling}.
BlueSpec has a well-define operational semantics and supports modular verification using SMT solvers~\cite{dave2011verification} and interactive-theorem provers~\cite{vijayaraghavan2015modular,choi2017kami}. 
While the use of high-level models helps raise the level of abstraction, and hence improves scalability in design and verification, 
all of these models including BlueSpec lacks two essential ILA features: a clean separation between hardware and software concerns, and uniform instruction-level treatment of processors and accelerators.
This limits their use in hardware/software co-verification and scalable verification of systems with heterogeneous hardware components. 

A number of hardware/software co-synthesis frameworks~\cite{legup-tecs-13,bambu-13,impluse-17,nane-16} attempt to automatically generate both firmware and accelerator hardware from an algorithmic description. While these efforts may 
side-step the need for abstractions for co-verification through correct-by-construction claims,  
reasoning about their correctness will itself require a principled abstraction of hardware with the key ILA features stated above.

Property validation of hardware over Verilog/VHDL models  has been advancing since the adoption of novel model checking techniques, e.g., \cite{jain2008word,vizel2012lazy,lee2014unbounded}. These works are orthogonal to our work.
Our key contribution is using ILA as a functional specification of processors/accelerators, and enabling the use of existing processor verification techniques for accelerator verification. The verification problem is to check equivalence of instruction-level vs. RTL models, and not validating individual properties in Verilog/VHDL models, which would otherwise need to be specified for capturing full functionality.
\section{Conclusions}
This paper presents the Instruction-Level Abstraction (ILA) as a formal model for accelerators to address the heterogeneity challenges of emerging computing platforms. The ILA is a uniform model, usable across heterogeneous processors and accelerators. Further, it raises the level of abstraction of the accelerators to that of the processors, enabling formal software-hardware co-verification. The ILA has several valuable  attributes for modeling and verification. It is modular, with functionality expressed as a set of instructions. It enables meaningful abstraction through architectural state that is persistent across instructions. It provides for portability through a more durable interface with the interacting processors.
It is hierarchical, providing for multiple levels of abstraction for modeling complex instructions as a software program through sub- and micro-instructions. 
It enables leveraging processor verification techniques for verifying accelerator implementations.
This allows for accelerator replacement using the notion of ILA compatibility similar to that of ISA compatibility.

We demonstrate the value of these attributes through modeling and verification of a range of accelerators (RBM, AES and Gaussian Blur) and a processor (RISC-V Rocket processor core). We identify modeling gaps in previous formal modeling of ISAs (ISA Formal's lack of distinction between hierarchy in specification vs. implementation) and a bug in the implementation of the RISC-V Rocket core. 
Further, we demonstrate substantially complete model checking based verification for our case studies.
Regarding scalability, our verification for accelerators from OpenCores (AES) and processors (Rocket Chip) are the targets over the next four years in the current DARPA POSH BAA.
Finally, we highlight additional applications of the ILA model in reasoning about concurrency and memory consistency with accelerators, accelerator code generation, and reliable simulator generation.
Overall, these results and contributions provide significant evidence of the value of ILAs in accelerator-based modeling and verification. 

\bibliographystyle{ACM-Reference-Format}
\bibliography{non-paper-reference,library-todaes}

\end{document}